\documentclass[a4paper]{jpconf}
\usepackage{graphicx,braket}
\begin{document}

\bibliographystyle{iopart-num}

\title{Signatures of octupole shape phase transitions in radioactive nuclei}
\author{Kosuke Nomura}
\address{Department of Physics, Faculty of Science, University of Zagreb, 10000 Zagreb, Croatia}
\ead{knomura@phy.hr}

\begin{abstract}
We analyze the octupole deformations and the related collective excitations in 
medium-heavy and heavy nuclei based on the microscopic framework of the 
nuclear energy density functional theory. Constrained 
self-consistent mean-field calculation with a given energy density functional 
is performed to provide for each nucleus 
a potential energy surface with axial quadrupole and octupole shape degrees 
of freedom. Spectroscopic properties are computed by means of the interacting-boson Hamiltonian, 
which is determined by mapping the fermionic potential energy surface onto the bosonic counterpart. 
The overall systematics of the 
calculated spectroscopic observables exhibit phase transitional behaviors 
between stable octupole deformation and octupole vibration characteristic of the 
octupole-soft potential within the set of nuclei in 
light actinide and rare-earth regions, Th, Ra, Sm, Gd, and Ba isotopes, where 
octupole shapes are most likely to occur. 
\end{abstract}

\section{Introduction}

The octupole (or pear-like) deformation in nuclei is one of the most prominent and studied 
themes in nuclear structure physics \cite{butler1996}. 
Measurement of permanent octupole deformation has an implication for new physics 
beyond the Standard Model of elementary particles. 
Experiments using radioactive-ion beams 
are planned or already operational around the world to find 
evidence for strong octupole deformation in several mass regions, 
e.g., $A\approx 220$ and $A\approx 144$.  
In this context, timely systematic and reliable nuclear structure calculations on 
octupole deformations and the related spectroscopic properties over a 
wide range of the chart of nuclides are required. 
We have carried out large-scale spectroscopic studies on 
octupole shapes and excitations in medium-heavy and heavy nuclei 
\cite{nomura2013oct,nomura2014,nomura2015,nomura2018oct}. 
The theoretical method is based on microscopic framework provided by the 
nuclear energy density functional theory (DFT) \cite{bender2003} 
and the interacting boson model (IBM) \cite{IBM}. 
The self-consistent mean-field (SCMF) calculation with a given  
relativistic or non-relativistic energy density functional (EDF) is performed to obtain 
potential energy surface (PES) in terms of the axial quadrupole $\beta_{20}$ and 
octupole $\beta_{30}$ shape degrees of freedom.  
The low-lying positive- ($\pi=+1$) and negative- ($\pi=-1$) parity states, and electromagnetic 
transition rates that characterize the octupole collectivity are computed by means of 
the IBM: The strength parameters of the IBM Hamiltonian, which comprises both 
positive- and negative-parity bosons, are completely determined by mapping the DFT 
energy surface onto the expectation value of the Hamiltonian in the boson condensate state. 
In this contribution we discuss specifically 
the quantum phase transitions (QPTs) of nuclear shapes \cite{cejnar2010}
with quadrupole and octupole degrees of freedom in Th, Ra, Sm, Gd, and Ba isotopes 
\cite{nomura2013oct,nomura2014,nomura2015}, 
and the octupole correlations in neutron-rich odd-mass Ba isotopes \cite{nomura2018oct}.

\section{Octupole shape phase transitions in light actinide and rare-earth nuclei}

\begin{figure}[h]
\begin{center}
\includegraphics[width=0.48\linewidth]{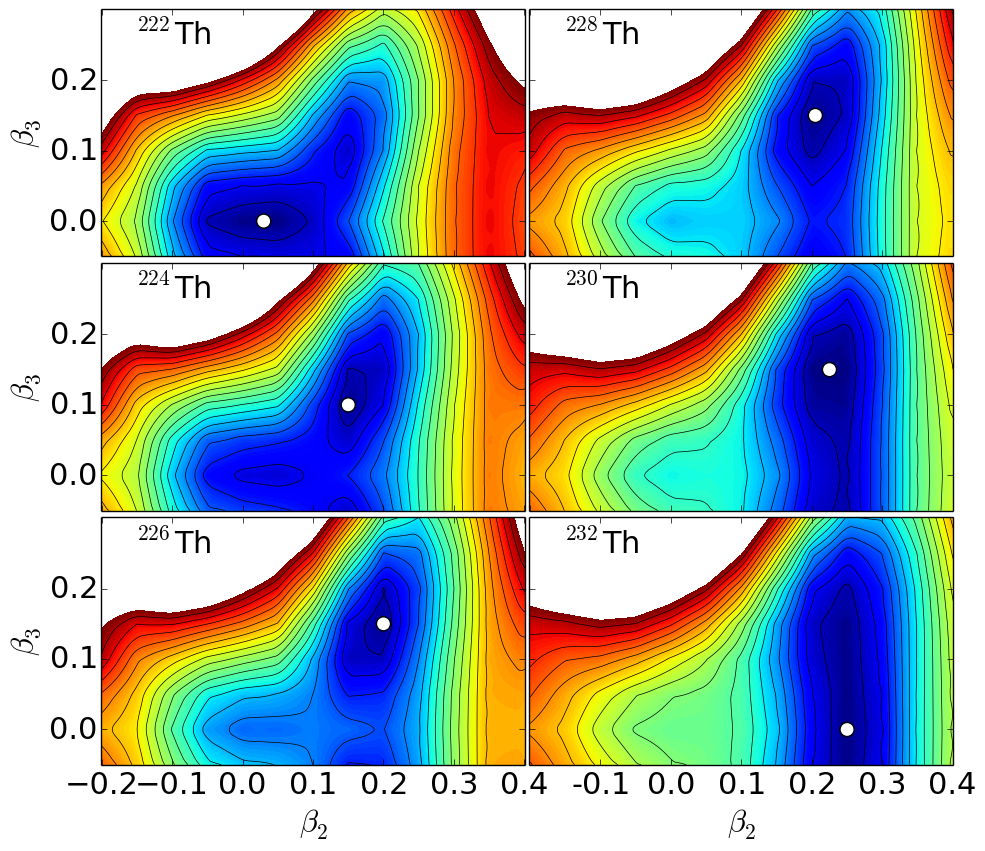} 
\includegraphics[width=0.48\linewidth]{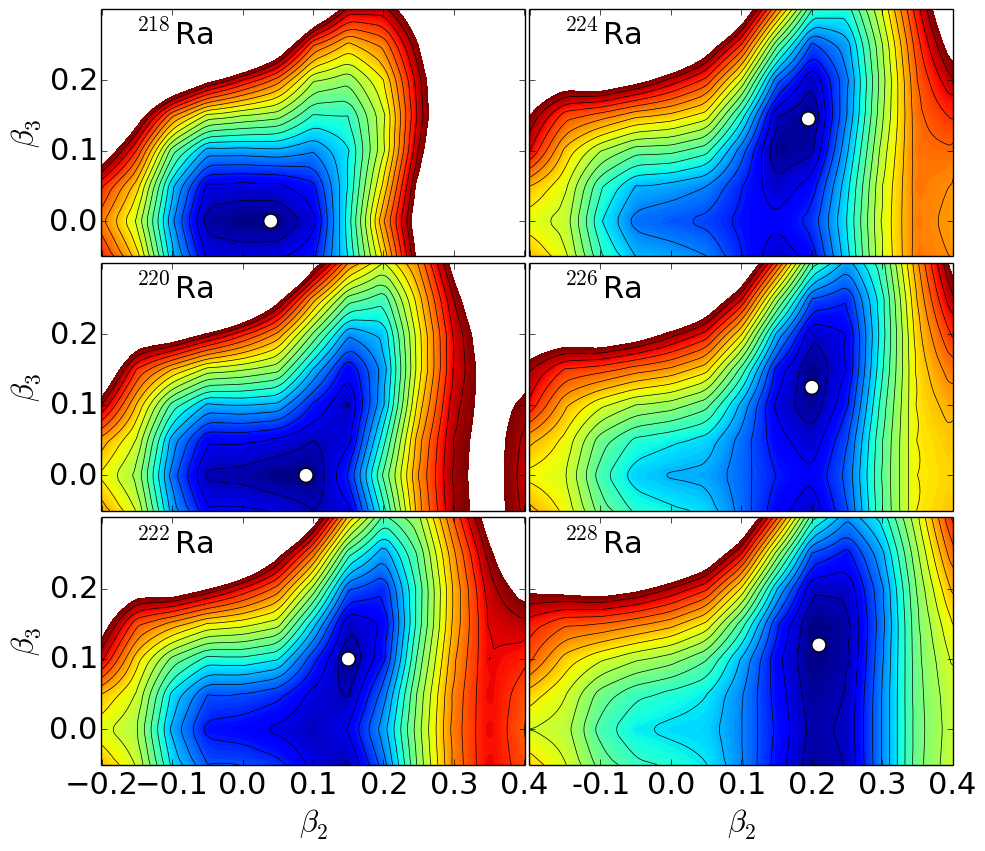}\\
\end{center}
\caption{\label{fig:pes-thra} Axially-symmetric $(\beta_{20},\beta_{30})$ SCMF PESs for the $^{222-232}$Th, 
and $^{222-232}$Ra isotopes. Contours are plotted with steps of 0.5 MeV, and the global minimum is identified by open circle.}
\end{figure}

\begin{figure}[h]
\begin{center}
\includegraphics[width=0.48\linewidth]{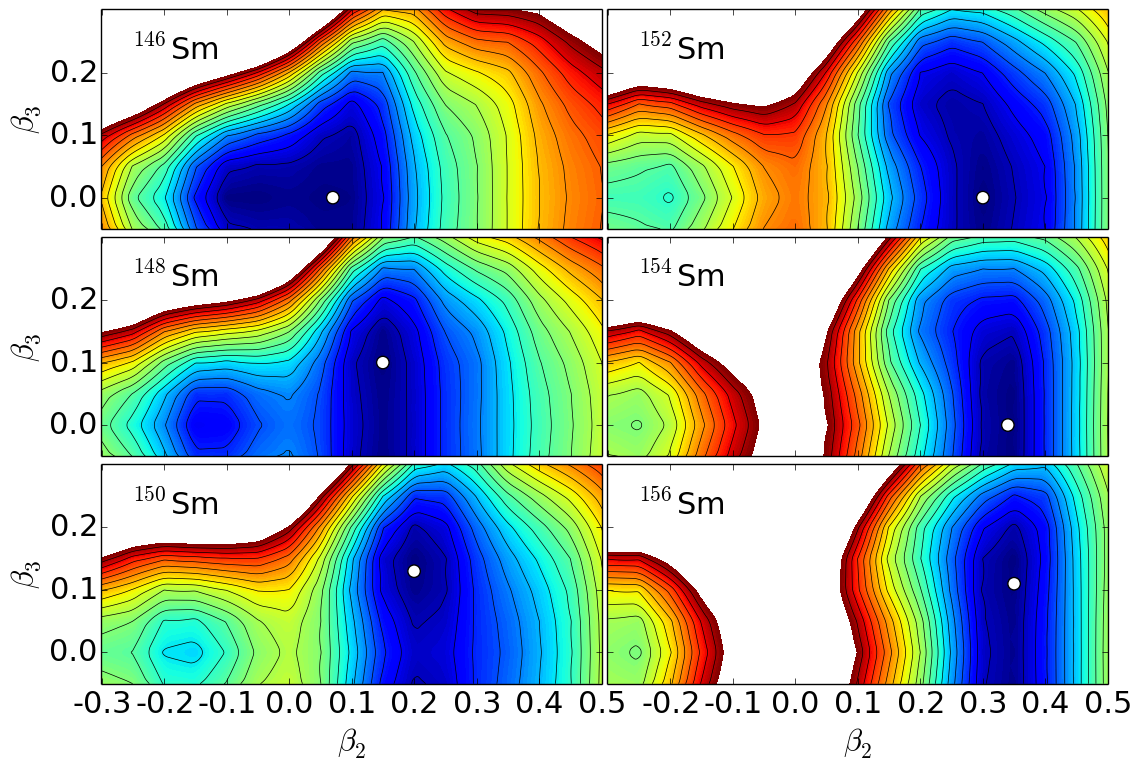} 
\includegraphics[width=0.48\linewidth]{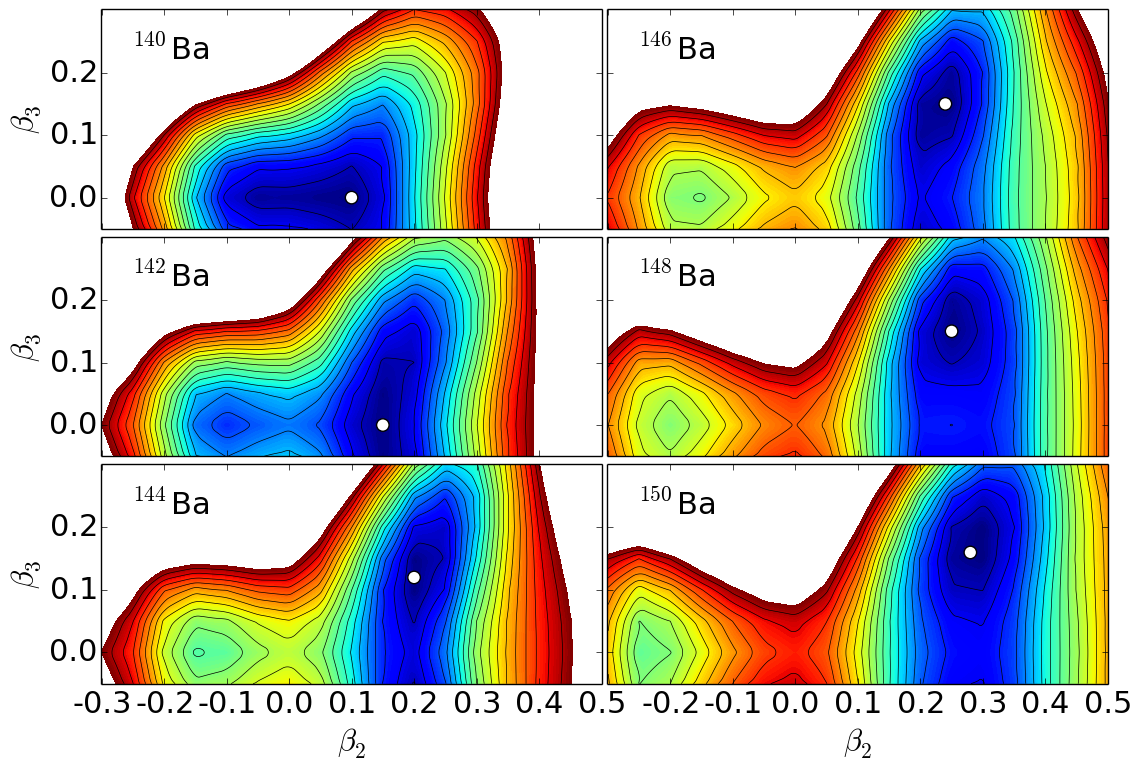}\\
\end{center}
\caption{\label{fig:pes-smba} Same as Fig.~\ref{fig:pes-thra}, but for the $^{222-232}$Sm, and $^{142-150}$Ba isotopes.}
\end{figure}

The axially-symmetric quadrupole and octupole PESs are shown in 
Figs.~\ref{fig:pes-thra} and \ref{fig:pes-smba}, that are computed by the 
constrained SCMF calculation within the relativistic Hartree-Bogoliubov method 
with the DD-PC1 EDF \cite{niksic2011}. 
Already at the SCMF level, features of the shape-phase transitions are observed: 
Non-zero $\beta_{30}$ deformation appears already at $^{224}$Th, 
and this octupole minimum becomes 
much more pronounced at $^{226,228}$Th, for which rigid octupole 
deformation is predicted. 
One then sees a transition to octupole-soft shapes at $^{230,232}$Th. 
The quadrupole $\beta_{20}$ deformation 
stays constant at $\beta_{20}\approx 0.2$ for $A\geq 226$. 
A similar observation applies to the PESs for the $^{220-230}$Ra isotopes. 
As for Sm isotopes (in Fig.~\ref{fig:pes-smba}), 
the most pronounced octupole minimum appears at around the neutron number 
$N=88$ ($^{150}$Sm) and, for heavier Sm isotopes, the octupole minimum is no longer present. 
Somewhat similar systematic is found for Ba.

For a more quantitative analysis of the QPT, it is necessary to compute 
spectroscopic properties, including excitation spectra and transition rates, 
by taking into account 
dynamical correlations beyond the mean-field approximation, i.e., those arising from 
symmetry restoration and fluctuations for the collective coordinates.  
To this end, we resort to the diagonalization of the IBM Hamiltonian, which is 
determined by mapping, at each configuration 
$(\beta_{20},\beta_{30})$ the SCMF PES 
$E_\mathrm{SCMF}(\beta_{20},\beta_{30})$, onto the bosonic one 
$E_\mathrm{IBM}(\beta_{20},\beta_{30})$, i.e., 
$E_\mathrm{SCMF}(\beta_{20},\beta_{30})\approx E_\mathrm{IBM}(\beta_{20},\beta_{30})$. 
The boson system consists of the positive-parity $0^+$ ($s$) and $2^+$ ($d$) 
and negative-parity $3^-$ ($f$) bosons. 
The bosonic PES is represented by the expectation value of the $sdf$-IBM Hamiltonian 
in the boson coherent state. 
See Refs.~\cite{nomura2008,nomura2014} for details of the whole procedure. 


\begin{figure}[h]
\begin{center}
\includegraphics[width=0.8\linewidth]{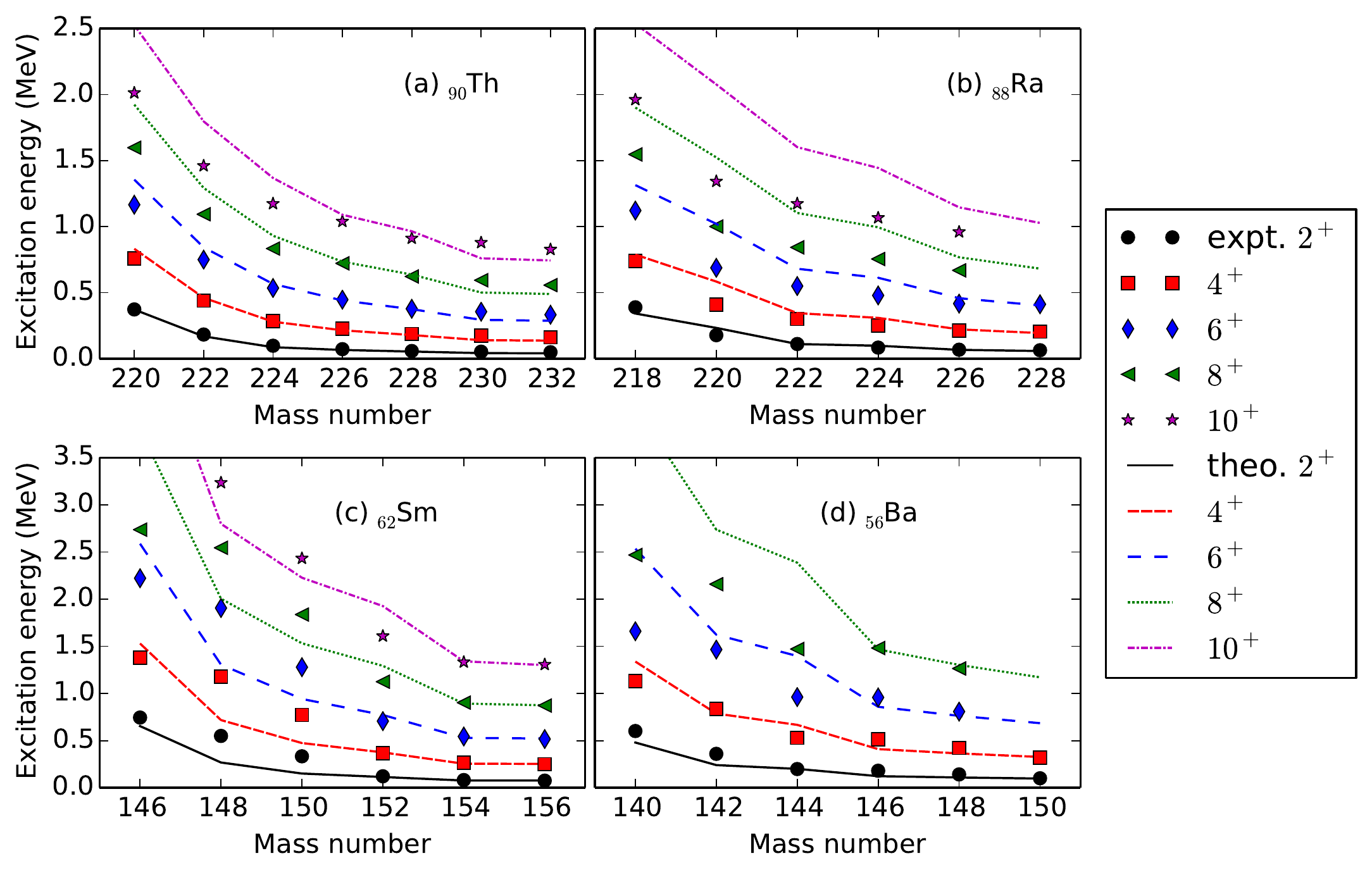}\\
\includegraphics[width=0.8\linewidth]{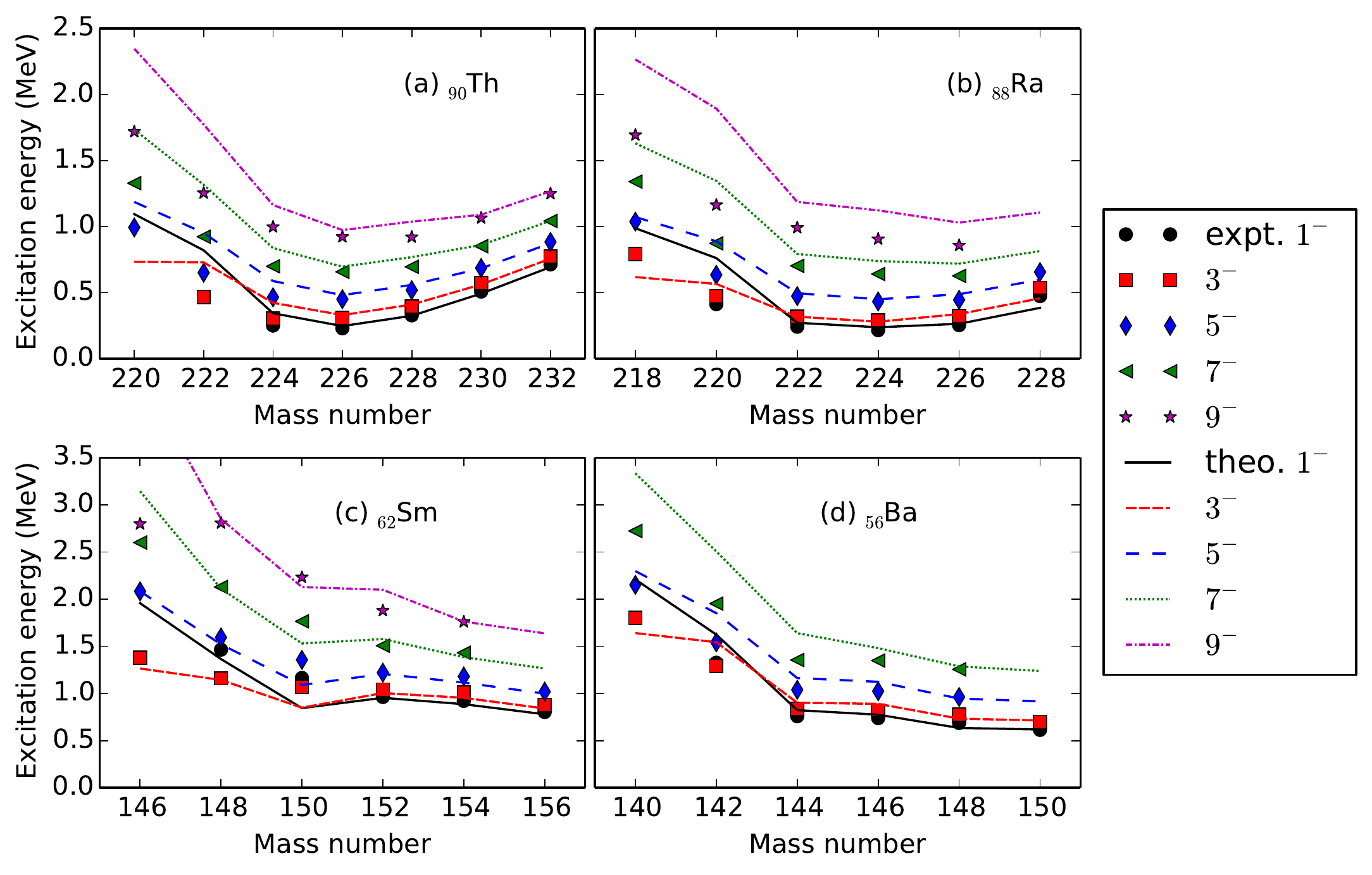}
\end{center}
\caption{\label{fig:level} 
Excitation energies for the $\pi=\pm 1$ yrast states in the 
Th, Ra, Sm, and Ba isotopes. }
\end{figure}

Figure~\ref{fig:level} depicts the calculated positive- and negative-parity low-lying levels in comparison with 
the experimental data. Firstly one should notice a very nice agreement between our 
calculation and the data, even though no phenomenological adjustment of the IBM 
parameters is made. In all the considered isotopic chains, the positive-parity yrast levels become lowered with increasing 
neutron number within each isotopic chain, suggesting spherical vibrational to 
strongly axially deformed states. What is of particular interest is the behaviors of the 
low-lying negative-parity states. They demonstrate a parabolic systematic as functions 
of the neutron number, cantered at a particular nucleus, e.g., $^{226}$Th, where 
the corresponding PES indicates the most pronounced 
octupole global minimum. In the Th isotopic chain, for instance, at $^{226}$Th the positive- 
and negative-parity bands are so close in energy to each other and seem to form 
an approximate alternating parity band typical of the stable octupole deformation. 
For those nuclei heavier than $^{226}$Th, however, both the positive- and 
negative-parity band begin to form separate bands. 
A similar result is obtained in the Ra isotopic chain. 
As for Sm, the negative-parity bands become lower in energy toward 
$^{150}$Sm but, from this nucleus on, stays rather constant, which means 
there is no notable change in the evolution of octupole collectivity. 
In Ba, the negative-parity levels become lowest at $^{144}$Ba and remain 
constant for the heavier Ba nuclei. 

\begin{figure}[h]
\begin{center}
\includegraphics[width=\linewidth]{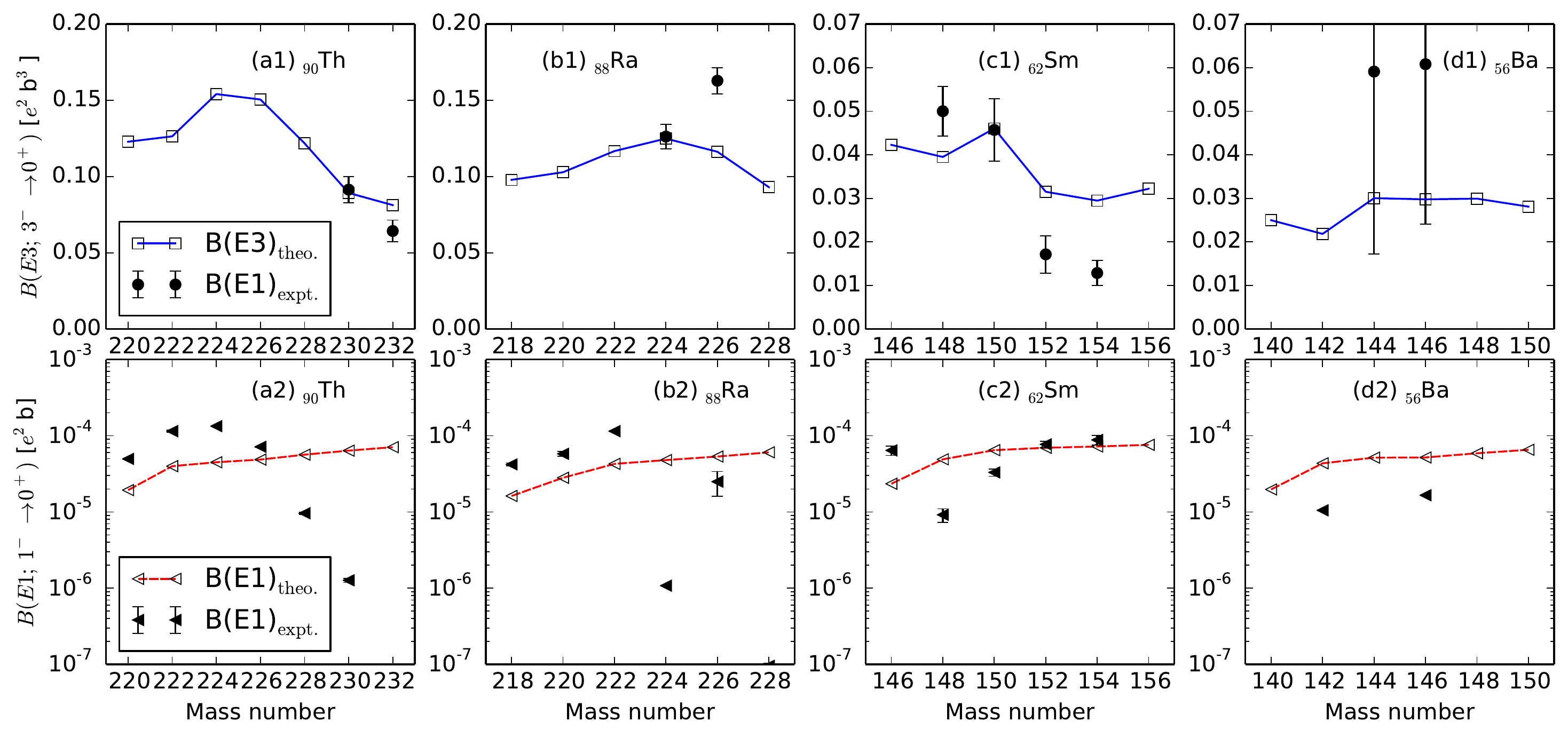}
\caption{\label{fig:trans} 
The $B(E3)$ and $B(E1)$ values for the considered Th, Ra, Sm, and Ba isotopes.}
\end{center}
\end{figure}

Next we show in Fig.~\ref{fig:trans}
the $B(E3; 3^-_1\to 0^+_1)$ and $B(E1; 1_1^-\to 0^+_1)$ transition rates. 
In particular the $B(E3)$ rates are a good 
measure for the octupole collectivity and, indeed, the predicted $B(E3)$ value 
becomes maximal at that nucleus 
where the PES exhibits the most pronounced 
$\beta_3\neq 0$ octupole minimum in each isotopic chain. 
On the other hand, the E1 property is accounted for 
by single-particle degrees of freedom, which is, by construction, 
not included in the model, as it is build only on the collective valence nucleons. 
That is the reason why the calculation fails to reproduce some experimental $B(E1)$ 
systematics.

\begin{figure}[h]
\begin{center}
\includegraphics[width=0.48\linewidth]{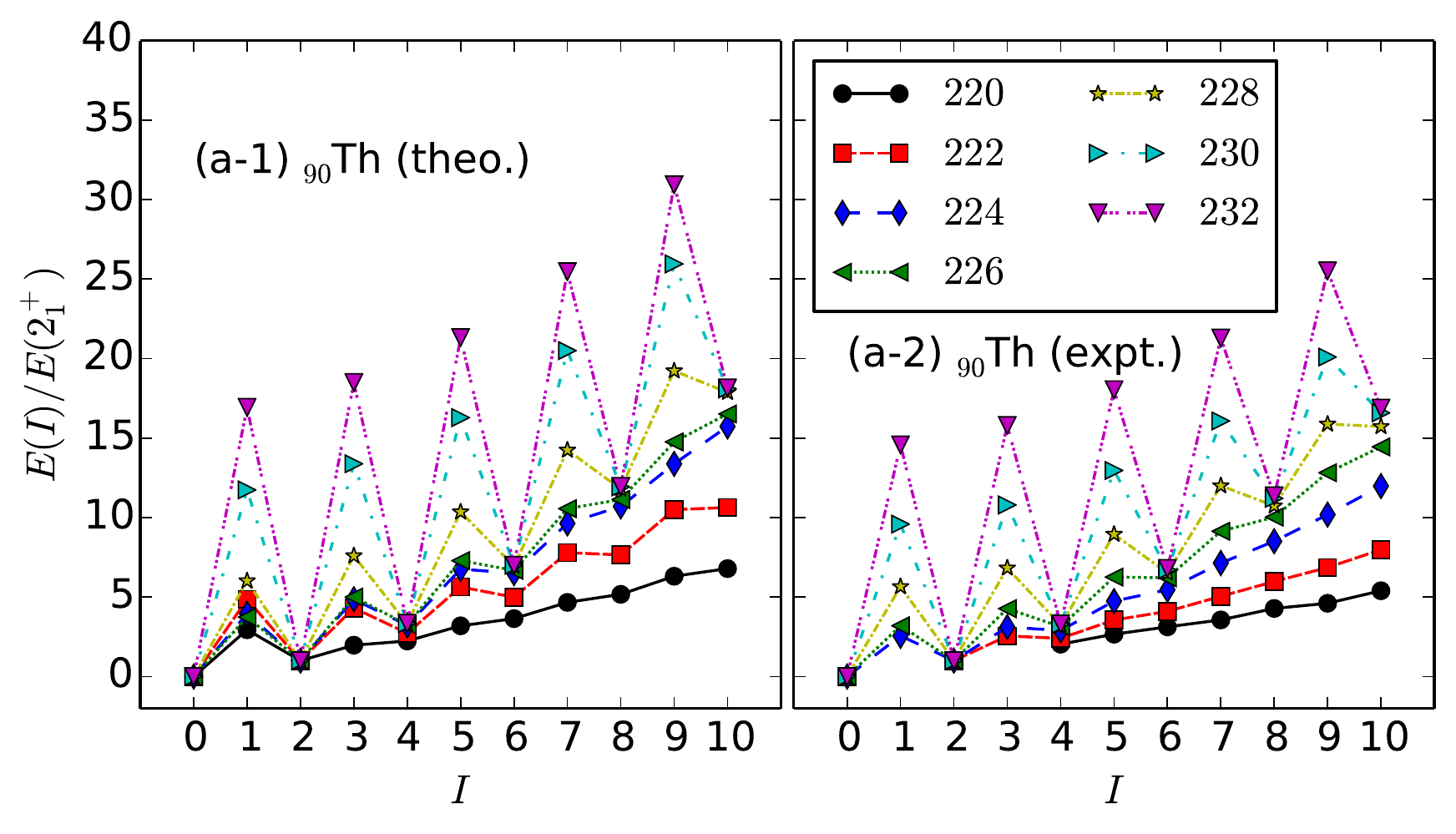} 
\includegraphics[width=0.48\linewidth]{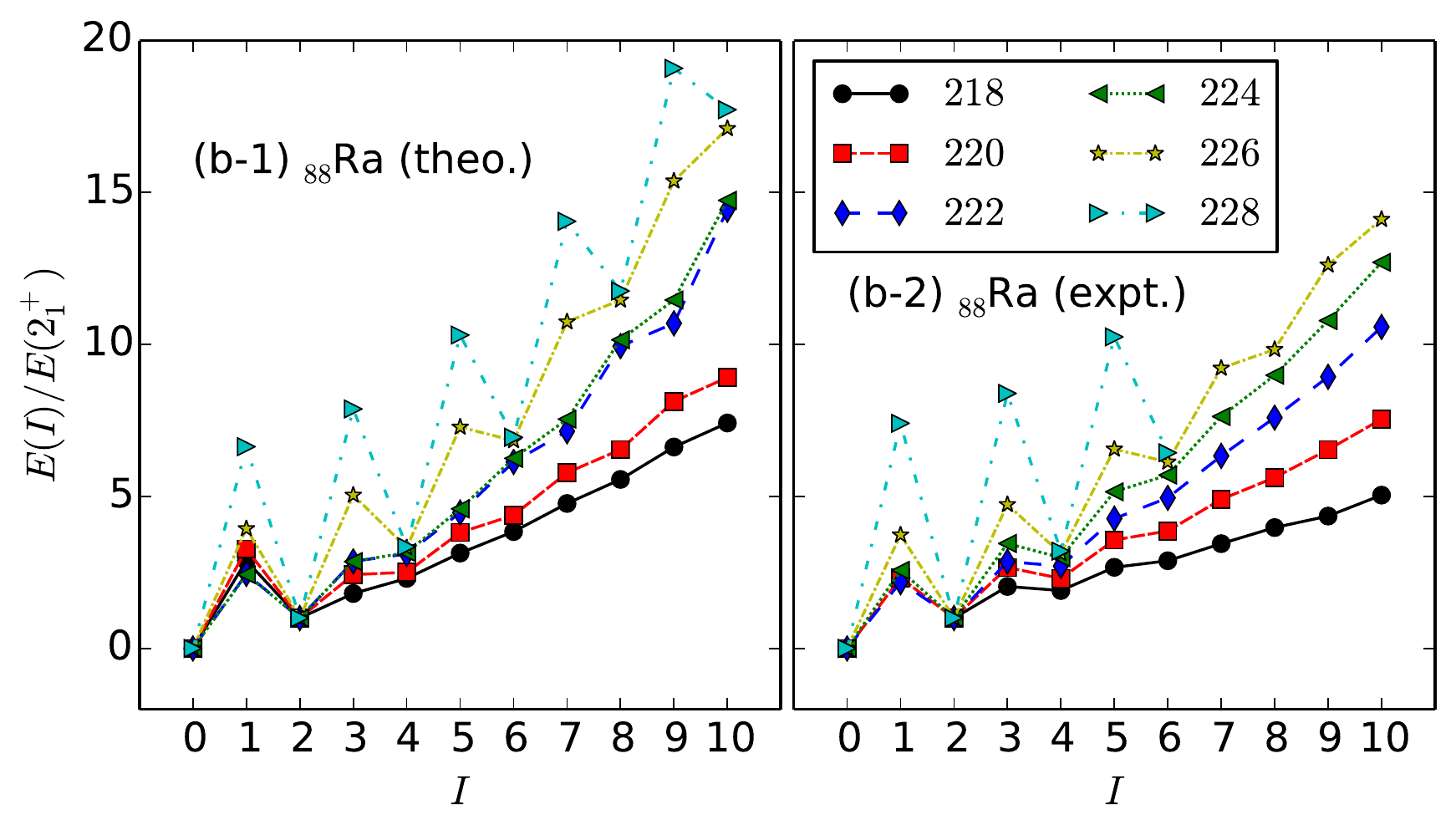} \\
\includegraphics[width=0.48\linewidth]{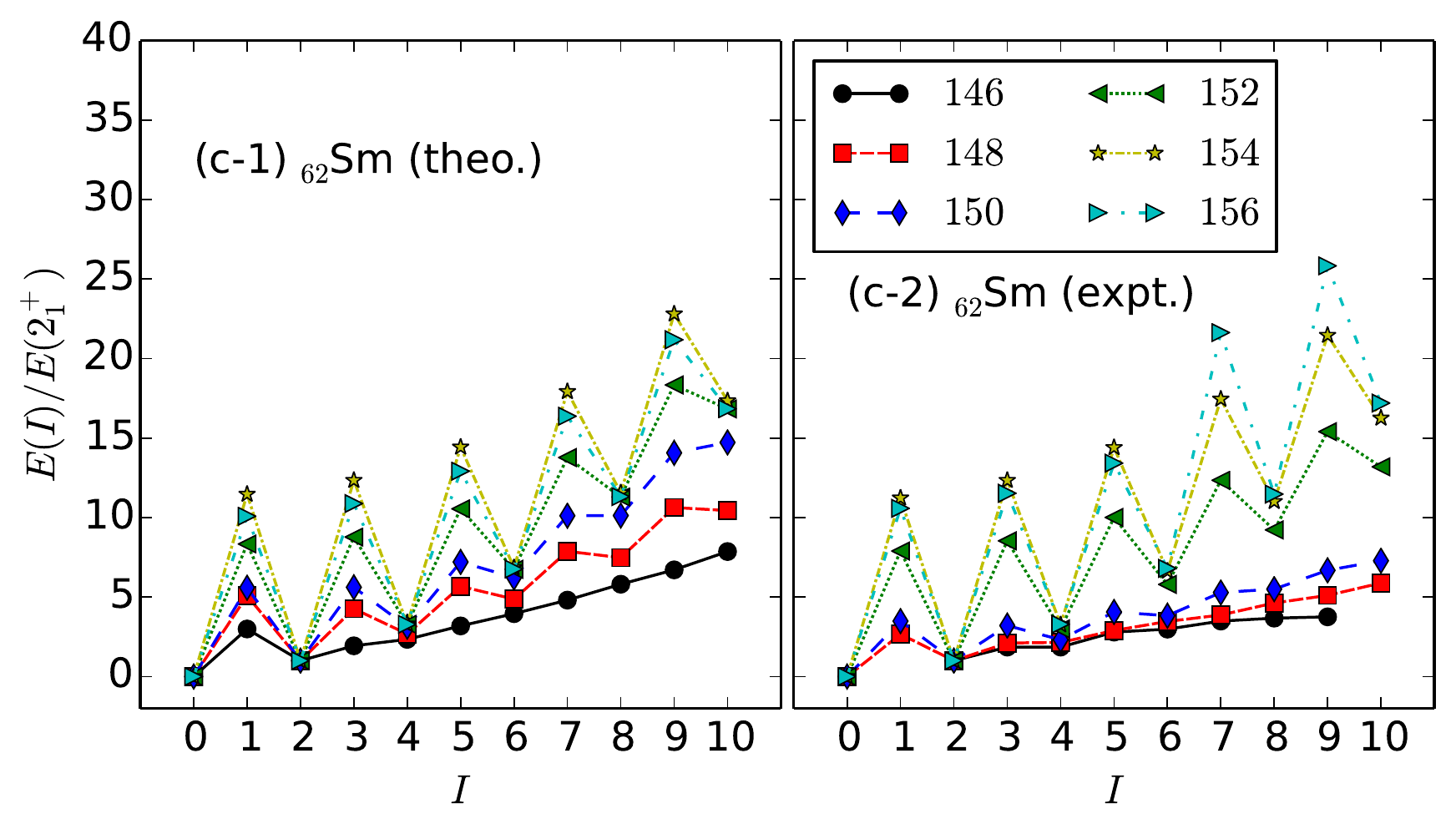} 
\includegraphics[width=0.48\linewidth]{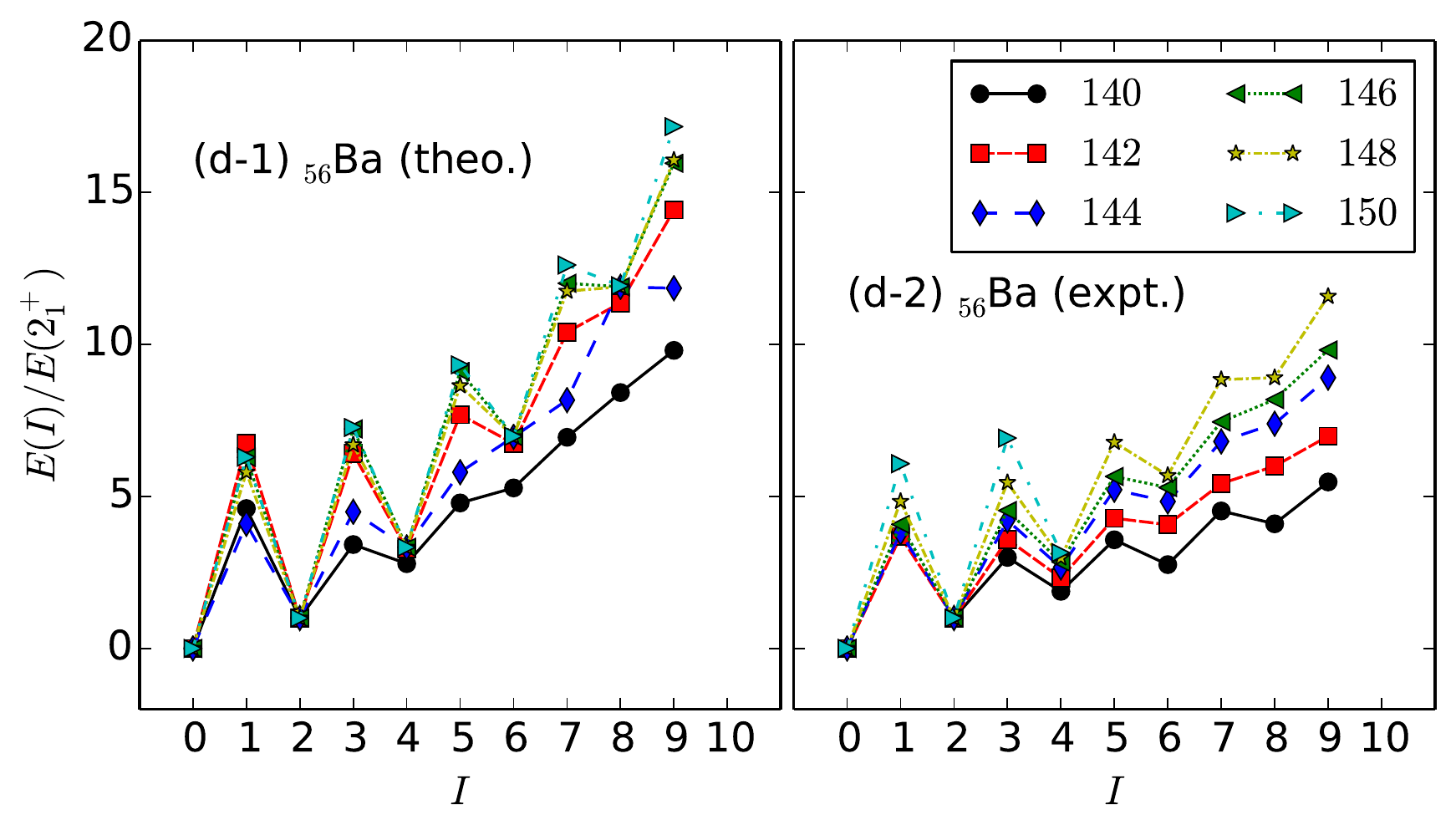} \\
\end{center}
\caption{\label{fig:ratio} Energy ratios $E(I^\pi)/E(2^+_1)$ for the $\pi=\pm 1$ yrast states with angular momentum $I$.}
\end{figure}

As another signature of the octupole QPT, we show in Fig.~\ref{fig:ratio} the energy ratio 
$E(I^\pi)/E(2^+_1)$ for the $\pi=\pm 1$ yrast states plotted against the 
angular momentum $I$. If the nucleus has stable octupole deformation 
and exhibits alternating-parity band, the ratio increases linearly with $I$. 
The staggering pattern shown in the figure, that starts from a particular nucleus, e.g., $^{226}$Th 
in the Th chain, indicates that the $\pi=+1$ and $\pi=-1$ yrast bands are decoupled and 
the octupole vibrational structure emerges.

We have also done a spectroscopic study on the octupole 
deformations in Sm and Gd isotopic chains by using the non-relativistic, Gogny EDF \cite{nomura2015}. 
There we confirmed the robustness of the mapping procedure:  irrespectively of whether 
relativistic or non-relativistic EDF is employed, a very nice description of the experimental 
low-lying positive- and negative-parity spectra, as well as the evolution of octupole deformation 
is obtained. 
Another interesting result in Ref.~\cite{nomura2015} is that many of the excited $0^+$ 
states in the considered Sm and Gd nuclei could have in their wave functions 
double octupole phonon (i.e., $f$ boson) component, and this result gives a possible 
explanation for why so many low-lying excited $0^+$ states are observed in rare-earth nuclei.

\section{Octupole correlations in odd-mass systems}

Extension to odd-mass system is made by introducing an unpaired nucleon, 
which is then coupled to the octupole deformed even-even nucleus as a core. 
The low-lying structure of even-even nucleus is described in terms of the interacting 
$s$, $d$, and $f$ bosons, and the particle-boson 
coupling is modeled within the interacting boson-fermion model (IBFM) \cite{IBFM}. 
The Hamiltonian for the IBFM consists of the $sdf$-IBM Hamiltonian $\hat H_\mathrm{B}$, 
the Hamiltonian for the single neutron $\hat H_\mathrm{F}$, and the term $\hat H_\mathrm{BF}$
that couples the fermion and boson spaces \cite{nomura2018oct}: 
$\hat H_\mathrm{IBFM} = \hat H_\mathrm{B}+\hat H_\mathrm{F}+\hat H_\mathrm{BF}$
%
Input from the SCMF calculation are the spherical single-particle energies $\epsilon_j$ 
(needed for $\hat H_\mathrm{F}$) and occupation numbers $v^2_j$ 
(for $\hat H_\mathrm{BF}$) for the odd particle in orbital $j$. 

\begin{figure}[h]
\begin{center}
\begin{tabular}{cc}
\includegraphics[width=0.48\linewidth]{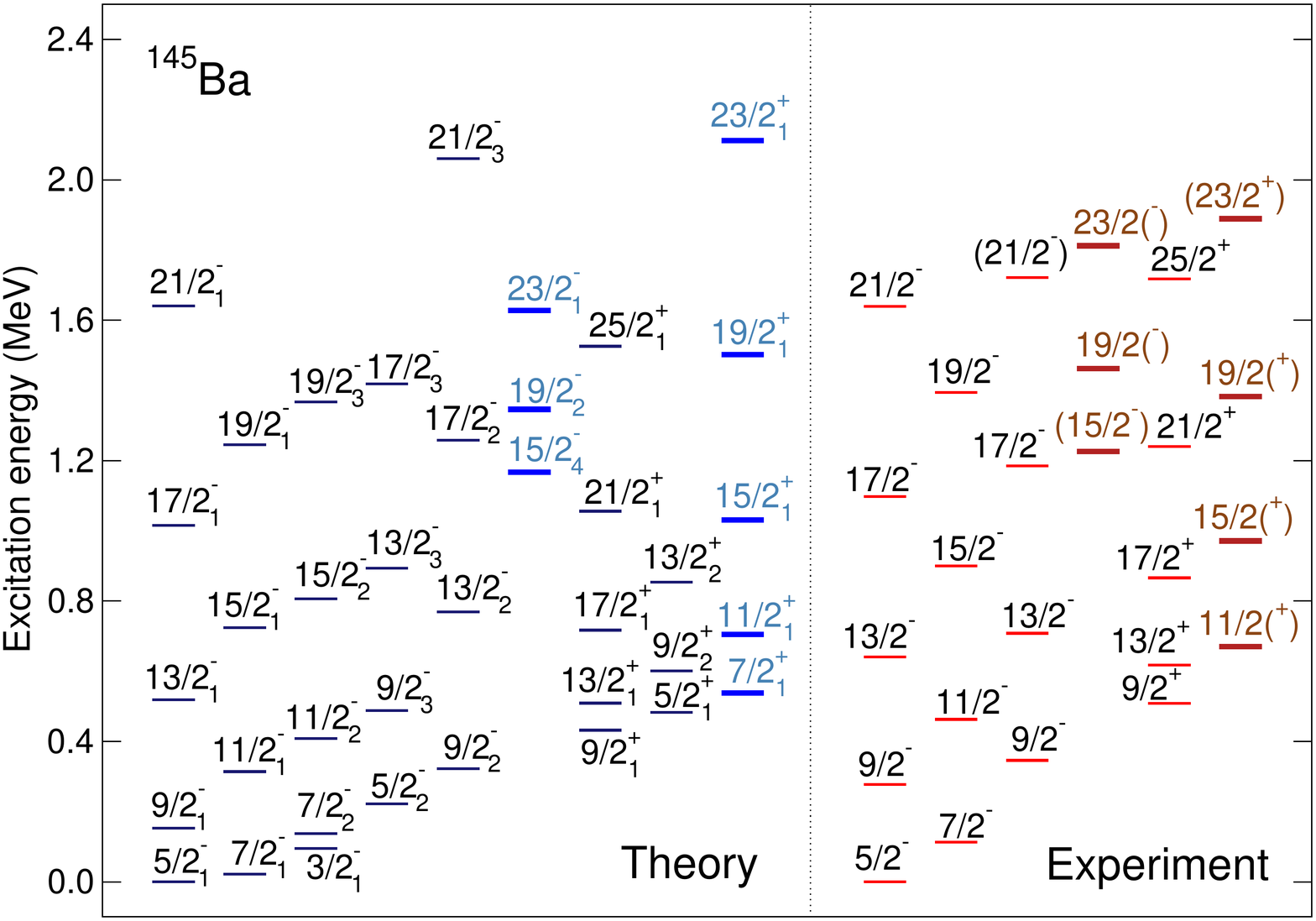} &
\includegraphics[width=0.48\linewidth]{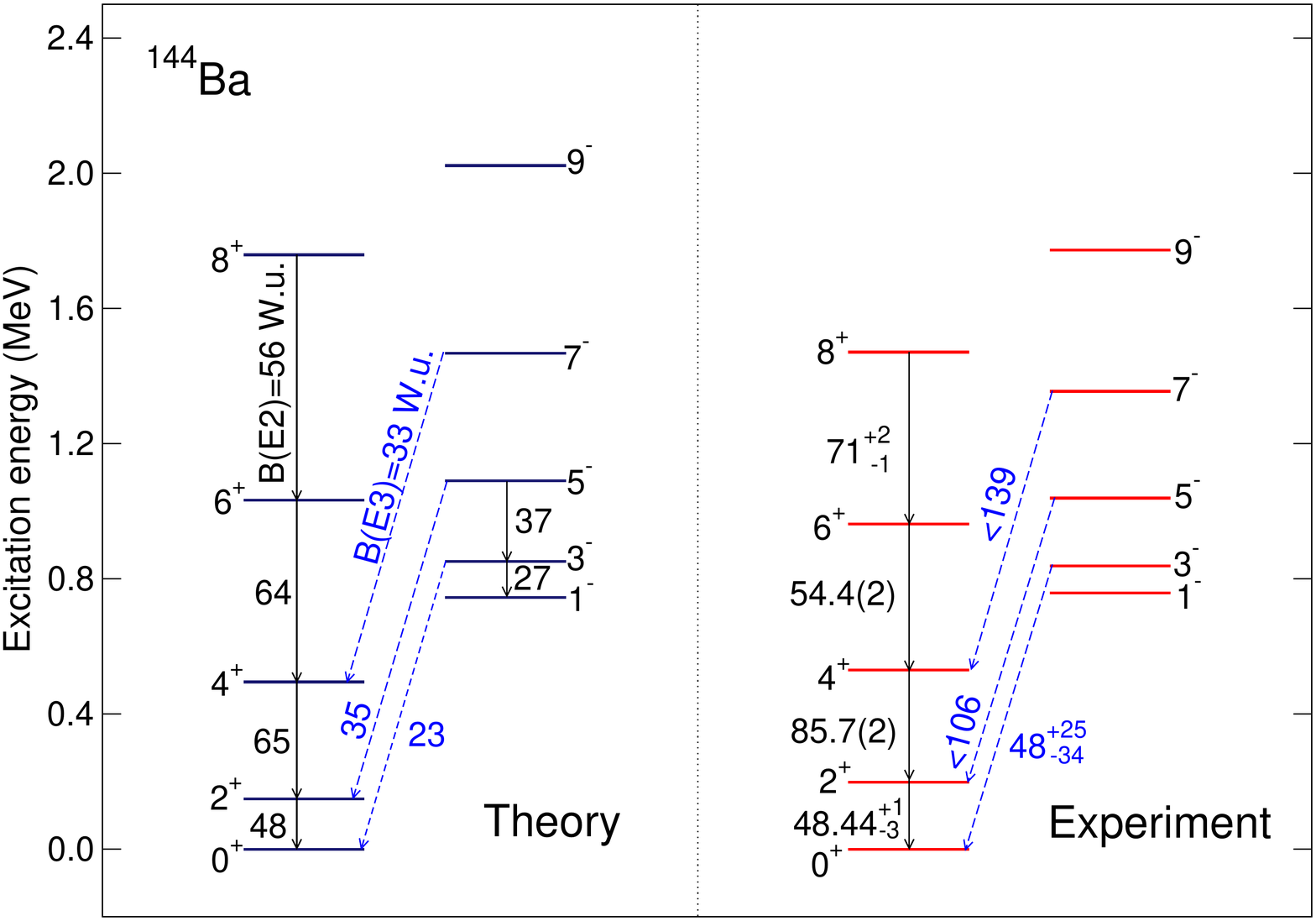}\\
\end{tabular}
\end{center}
\caption{\label{fig:ba}The theoretical and experimental excitation spectra for the $^{145}$Ba and $^{144}$Ba nuclei.}
\end{figure}

Here we illustrate the method in its application to the isotope $^{145}$Ba. 
Since the corresponding even-even boson core nucleus $^{144}$Ba exhibits 
an octupole-soft potential at the SCMF level, we also expect that the octupole correlations play 
an important role in the low-energy spectra of this odd-mass nucleus. 
The calculated excitation spectrum for $^{145}$Ba is compared to the 
corresponding experimental bands in Fig~\ref{fig:ba}. 
Those calculated positive- and negative-parity bands shown 
in bold in the figure are made of the 
one-$f$-boson configuration coupled with a single neutron in 
the  $p_{1/2,3/2}f_{5/2,7/2}h_{9/2}$ and $i_{13/2}$ orbitals, respectively. 
The corresponding experimental bands that are 
suggested to be of octupole in nature are also depicted in the figure. 
The absolute energies of the bandheads and energy spacings 
within the bands are well reproduced by the calculation. 
We have also predicted the $B(E3)$ transition rates from the octupole bands to the 
ground-state bands to be typically within the range 20$\sim$30 W.u., which are 
of the same order of magnitude as the calculated $B(E3;3^-\rightarrow 0^+)$ value of 23 W.u. 
for the neighboring even-even nucleus $^{144}$Ba. To examine quality of the 
model prediction, more experimental information about the $B(E3)$ transitions 
for the odd-mass Ba isotopes is expected.

\section{Summary}
Based on a global theoretical framework of the nuclear DFT, we have analyzed the 
octupole deformations and the related spectroscopic properties 
in a large set of medium-heavy and heavy nuclei. The results of the SCMF 
calculations with a given relativistic and non-relativistic EDF are used to completely 
determine the Hamiltonian of the IBM, which then provides excitation spectra and transition rates. 
Evolution of the axially-symmetric quadrupole and octupole PES, 
the resultant positive- and negative-parity low-lying states, and E3 transition rates 
and moments, all consistently points to the onset of octupole deformations and the 
phase transition between stable octupole deformation and octupole soft shapes in 
the considered Th, Ra, Sm, Gd, and Ba isotopes chain. 
The octupole correlation plays 
an important role in describing low-lying spectra in odd-mass nuclei 
as well as in neighboring octupole deformed even-even nuclei. 
The theoretical method presented here is general and allows a computationally 
feasible prediction of octupole collective states and will be, therefore, 
extended further to study many other radioactive nuclei that are becoming 
of much more importance for the RIB experiments. 

\section*{Acknowledgments}
The results presented in this contribution are based on the works with 
D. Vretenar, T. Nik{\v s}i\'c, B.-N. Lu, L. M. Robledo, and R. Rodr\'iguez-Guzm\'an,
This work is financed within the Tenure Track Pilot Programme of the 
Croatian Science Foundation and the 
\'Ecole Polytechnique F\'ed\'erale de Lausanne and the Project TTP-2018-07-3554
 Exotic Nuclear Structure and Dynamics, with funds of the Croatian-Swiss Research Programme.

\bibliography{iopart-num}

\end{document}